\documentclass[12pt,]{article}
\usepackage{authblk}
\usepackage{fullpage}
\usepackage{amsmath,amsfonts,amssymb}
\usepackage[utf8x]{inputenc}
\usepackage[T1]{fontenc}
\usepackage{siunitx}
\usepackage{comment}
\usepackage[version=3]{mhchem}
\usepackage{booktabs}
\usepackage{float}
\usepackage{algorithmic}
\usepackage{array}
\usepackage[caption=false,font=normalsize,labelfont=sf,textfont=sf]{subfig}
\usepackage{textcomp}
\usepackage{stfloats}
\usepackage{url}
\usepackage{verbatim}
\usepackage{graphicx}
\usepackage{cite}
\usepackage{balance}
\usepackage{xcolor}

\def\BibTeX{{\rm B\kern-.05em{\sc i\kern-.025em b}\kern-.08em
    T\kern-.1667em\lower.7ex\hbox{E}\kern-.125emX}}

\graphicspath{{figures/}}

\title{Good Practices for Measuring the Interfacial Dzyaloshinskii-Moriya Interaction by Brillouin Light Scattering and Magneto-Optical Kerr Effect}

\author[1]{Marco Madami} 
\author[2]{Silvia Tacchi} 
\author[3]{Adriano Di Pietro}
\author[3]{Alessandro Magni} 
\author[4]{Giovanni Gandini}
\author[4] {Christian Rinaldi}
\author[1]{Giovanni Carlotti}
\author[3]{Gianfranco Durin}  
\author[5] {Johannes Hertel}
\author[5] {Maximilian Lederer}
\author[6]{Christopher H. Marrows} 
\author[7]{Vito Puliafito}
\author[3]{Michaela Kuepferling}

\affil[1]{Dipartimento di Fisica e Geologia, Università di Perugia, Via Pascoli, 06123 Perugia, Italy}
\affil[2]{CNR, Istituto Officina dei Materiali - Perugia, c/o Dipartimento di Fisica e Geologia, Università di Perugia, Via A. Pascoli, 06123 Perugia, Italy}
\affil[3]{Istituto Nazionale di Ricerca Metrologica, Strada delle Cacce 91, 10135, Torino, Italy}
\affil[4]{PoliMi, Dipartimento di Fisica, Politecnico di Milano, 20133 Milano, Italy}
\affil[5]{Fraunhofer IPMS (Center Nanoelectronic Technologies CNT), 01109 Dresden, Germany}
\affil[6]{School of Physics and Astronomy, University of Leeds, Leeds LS2 9JT, United Kingdom}
\affil[7]{Department of Electrical and Information Engineering, Politecnico di Bari, 70126 Bari, Italy }

\begin{document}

\maketitle

\newpage

\begin{abstract}
The interfacial Dzyaloshinskii-Moriya interaction (iDMI) is a key ingredient for stabilizing chiral magnetic states, such as skyrmions and chiral domain walls, which are key building blocks for next-generation spintronic technologies. Despite its importance, the quantitative determination of the iDMI constant, $D$, is often plagued by significant discrepancies. This paper establishes a set of good practices for measuring iDMI using two primary techniques: Brillouin Light Scattering (BLS) and the Magneto-Optical Kerr Effect (MOKE). Through the analysis of Co/Pt-based heterostructures, this work provides a comprehensive roadmap for measurement procedures, data evaluation, and uncertainty analysis. Our analysis highlights the crucial role of magnetic parameters used to extract the iDMI constant and propagation of their uncertainties, and the measurement model for enhancing measurement reliability. The proposed methodology provides a practical framework for improving the consistency and reproducibility of iDMI measurements across different laboratories and experimental techniques.

\end{abstract}

\section{Introduction}
The development of next-generation spintronic technologies, such as racetrack memory and logic-in-memory architectures, relies  on the energy-efficient manipulation of chiral magnetic states like chiral domain walls (DWs) and skyrmions \cite{Dieny2020, Parkin2015}. These textures are particularly attractive as information carriers because electric currents can drive them with high efficiency. A fundamental requirement for stabilizing these topological spin structures is the Dzyaloshinskii-Moriya interaction (DMI).

Originally proposed in the 1950s to explain weak ferromagnetism in bulk antiferromagnets like $\alpha$-Fe$_{2}$O$_{3}$ \cite{Dzi1957, Moriya1960}, the DMI  is a key ingredient for non-collinear or helical magnetic states in various systems, including spin glasses, orthoferrites, manganites, and superconducting cuprates \cite{Fert1980, Luo1999, Bogdanov2002, Coffey1991}. While the DMI often cancels out in bulk materials due to crystal inversion symmetry, a finite contribution emerges when this symmetry is broken. In thin-film heterostructures, symmetry breaking occurs at the interface between a ferromagnet (FM) and a heavy metal (HM) with strong spin-orbit coupling \cite{Hellman2017}. In these systems, the interaction becomes interfacial in nature (iDMI), therefore its strength $D$ is inversely proportional to the magnetic layer thickness \cite{MOO-13, Hellman2017}. The iDMI is crucial for stabilizing Néel-type domain walls with a fixed chirality \cite{Heide2008, thiaville2012}, which are energetically favored over traditional Bloch walls and enable fast current-driven dynamics \cite{Emori2013, Ryu2013}. 

However, despite its technological importance, quantitatively determining the iDMI constant $D$ remains challenging. Recent literature, including our review \cite{soucaille2016, SHA-19, herrera_diez_nonvolatile_2019, Kuepferling2023}  and our international interlaboratory round-robin comparison conducted across European, US, and Asian metrology and research institutes \cite{magni2022}, reports notable discrepancies in $D$ values for nominally identical material systems, particularly when evaluated using different experimental techniques.

To address these inconsistencies, this work aims to establish  concrete, step-by-step measurement protocols and good practices that minimize uncertainty and maximize interlaboratory reproducibility for measuring iDMI. We focus on two of the most prevalent techniques in the field \cite{Kuepferling2023}, which were already employed in the round-robin comparison: Brillouin Light Scattering (BLS) and the Magneto-Optical Kerr Effect (MOKE). While BLS relies on the non-reciprocity of spin-wave dispersion \cite{MOO-13, UDV-09, kostylev_non-reciprocity_2013, cortes-ortuno_influence_2013, nembach_linear_2015, DI-15, MA-17, Lin2020}, MOKE characterizes iDMI through domain wall dynamics within the thermally activated creep regime \cite{lemerle1998, je2013, vanatka2015, hrabec2014, hartmann2019, DiPietro2025}.  While we leave the detailed discussion of quantitative discrepancies between the two methods \cite{magni2022} for future work, here we focus on overcoming common pitfalls in the measurement procedures, data evaluation and uncertainty analysis. Particular key roles are attributed to the magnetic field reversal for BLS and minimum bubble size for MOKE. Furthermore, we determine spatial iDMI statistics, assess their impact on uncertainty, and correlate the results with sample homogeneity, a factor often neglected due to the substantial laboratory time required. Overall, the proposed protocols ensure reliable and reproducible results for both methods and offer a practical roadmap toward the standardized characterization of iDMI in chiral magnetic systems.

\section{Determination of iDMI by Brillouin Light Scattering}

\subsection{Physical Principle: Spin-Wave Non-reciprocity}

Brillouin light scattering (BLS) is a widely used technique for probing spin-wave (SW) dispersion and non-reciprocity. In a typical BLS experiment, a monochromatic laser beam is focused onto the sample surface, and the backscattered light is collected by the same objective and analyzed in frequency by a high-resolution spectrometer. During the scattering process, magnons can be either created (Stokes process) or annihilated (anti-Stokes process), resulting in a redshift or blueshift of the scattered light relative to the incident beam. In wave-vector space, Stokes and anti-Stokes processes correspond to magnons with opposite wave vectors. Because BLS probes thermally populated spin waves, it provides a non-destructive method to investigate spin dynamics without requiring external microwave excitation.

Numerous studies over the past decade have determined the iDMI constant $D$ using BLS, particularly in heavy-metal/ferromagnet (HM/FM) systems typically consisting of Co or CoFeB in contact with heavy metals such as Pt, W, or Ta \cite{Kuepferling2023}. The method exploits the fact that, in the presence of iDMI, counterpropagating spin waves exhibit different frequencies—i.e., the dispersion becomes non-reciprocal. To measure iDMI in thin films or multilayers, the Damon–Eshbach (DE) geometry is typically employed. In this configuration, the film is magnetized in-plane by a sufficiently large external magnetic field $H$, while the spin-wave wave vector $k$ lies perpendicular to the magnetization. In the presence of iDMI, this geometry produces a non-reciprocal spin-wave dispersion. The magnitude of the SW wave vector is determined by the angle of incidence $\theta$ through conservation of in-plane momentum:
\begin{equation}
    k = 2 k_{\mathrm{i}}^{\mathrm{light}} \sin \theta
    \label{eq:BLS_k}
\end{equation}
The iDMI contribution introduces a linear frequency shift:
\begin{equation}
    f_{\mathrm{DMI}} = \frac{\gamma D}{\pi M_{\mathrm{s}}} k
    \label{eq:BLS_fDMI}
\end{equation}
where $\gamma$ is the gyromagnetic ratio and $M_{\mathrm{s}}$ is the saturation magnetization. Therefore, the frequency difference between counterpropagating spin waves is directly proportional to both the wave vector $k$ and the iDMI constant $D$.

\begin{figure}[!t]
    \centering
    \includegraphics[width=1.1\linewidth, page=1, trim={2cm 1.5cm 2cm 0.5cm}, clip]{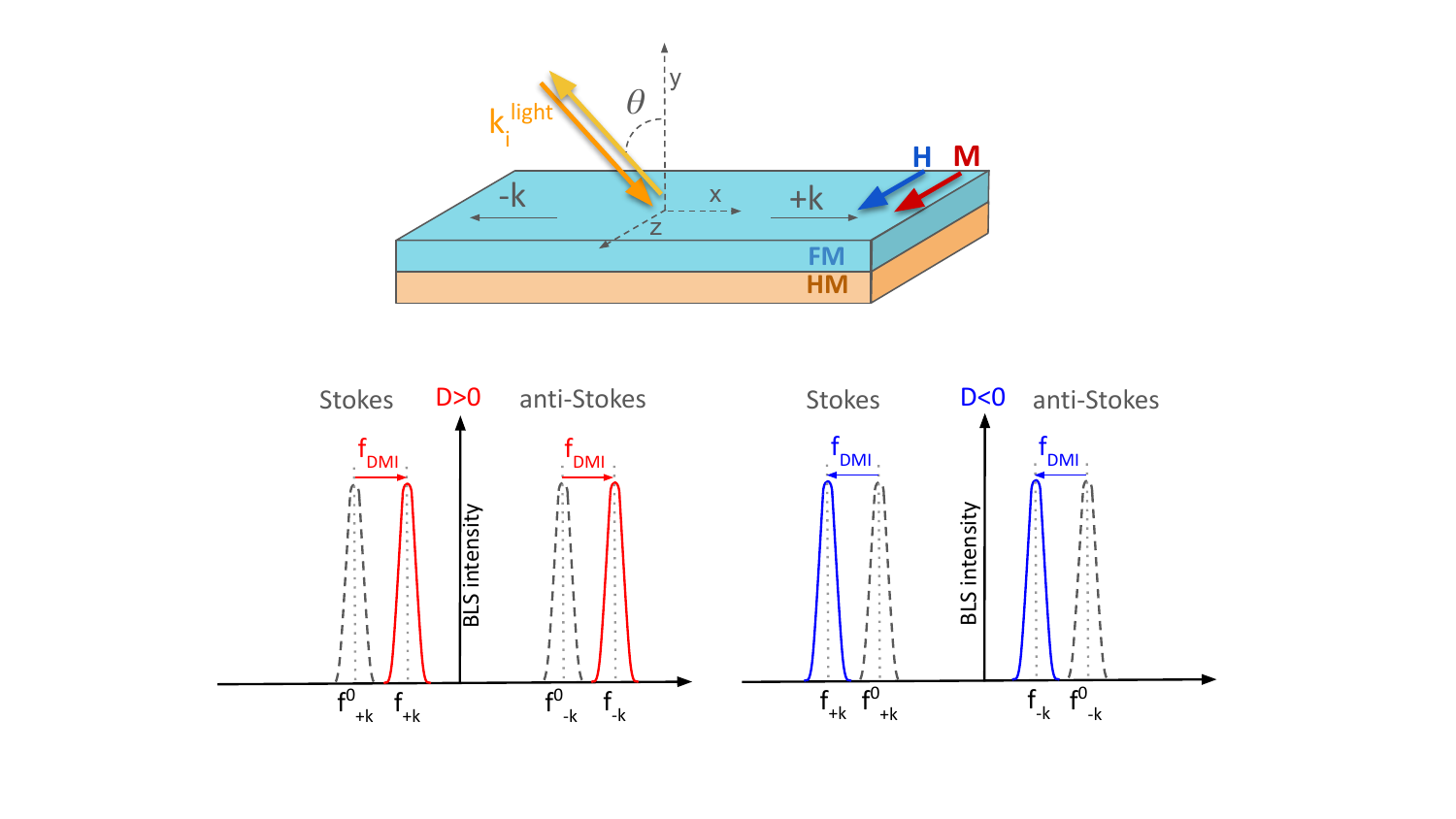}
    \caption{Schematic diagram of the BLS interaction geometry. The bottom panel shows a typical BLS spectrum in the absence of iDMI (dashed line) and with iDMI (solid line) for negative and positive values of $D$, respectively. Here $k_{\mathrm{i}}^{\mathrm{light}}$ denotes the incident light wave numbers, while $H$ and $M$ are the applied in-plane field and magnetization, respectively.}
    \label{fig:BLS_1}
\end{figure}

For an FM/HM bilayer, the sign of $D$ determines the favored magnetic chirality and depends on the Hamiltonian convention adopted; here, we use the plus-sign convention, where positive $D$ corresponds to Néel domain walls with right-handed (clockwise) chirality ($\uparrow \rightarrow \downarrow$ or $\downarrow \leftarrow \uparrow$), while negative $D$ corresponds to left-handed (anticlockwise) chirality ($\uparrow \leftarrow \downarrow$ or $\downarrow \rightarrow \uparrow$) \cite{Kuepferling2023}. As shown in the lower panels of Fig.~\ref{fig:BLS_1}, a negative (positive) value of $D$ produces a negative (positive) frequency shift $f_{\mathrm{DMI}}$, resulting in a rigid relative displacement of the Stokes and anti-Stokes peaks.

\subsection{Measurement Protocol}

Brillouin light scattering measurements were performed using a monochromatic light source ($\lambda = 532\text{ nm}$) provided by a single-mode diode-pumped solid-state laser. The optical path utilized a camera objective with a focal length of $50\text{ mm}$ and a numerical aperture of $0.25$. Analysis of the backscattered light was conducted via a Sandercock-type (3+3)-pass tandem Fabry–Perot interferometer. The magnitude of the in-plane wave vector $k$ is defined by the incidence angle $\theta$ and wavelength $\lambda$ through $k = \frac{4\pi \sin \theta}{\lambda}$ (here $k = 1.67 \times 10^7\text{ rad/m}$ at $\theta = 45^\circ$). The sample was mounted on a rotary non-magnetic stage between the poles of an electromagnet capable of generating fields up to $1.5\text{ T}$.

Reversing the direction of the applied magnetic field, and thus the sign of $f_{\mathrm{DMI}}$, is essential for obtaining a reliable value of $D$. The presence of iDMI is not the sole cause of a rigid frequency shift: even in a symmetric spectrum, minor optical alignment differences between the scattered light and the stabilization reference beam can introduce an instrumental asymmetry, typically on the order of $0.1\text{--}0.3\text{ GHz}$. In conventional BLS measurements, this effect is negligible because the spin-wave frequency can be extracted by averaging the Stokes and anti-Stokes peaks. However, when probing systems with iDMI, this physical asymmetry combines with the instrumental offset, biasing $f_{\mathrm{DMI}}$.

 In our experimental protocol, the two spectra under opposite field directions are acquired sequentially, immediately one after the other, on the exact same position on the sample. This back-to-back experimental protocol effectively minimizes any time-dependent thermal or mechanical drifts of the optical setup, making any temporal variation in the field-independent frequency offset negligible. Furthermore, because reversing the external magnetic field does not require any mechanical adjustment of the sample stage or optical alignment, it guarantees that the probed micro-scale volume remains identical for both field directions, preserving spatial reproducibility. The iDMI-induced frequency shift is then extracted as:
\begin{equation}
    f_{\mathrm{DMI}} = \frac{\left(f_{+k}(+H) + f_{-k}(+H)\right) - \left(f_{+k}(-H) + f_{-k}(-H)\right)}{4}
    \label{eq:BLS_2spectra}
\end{equation}
The uncertainty in $f_{\mathrm{DMI}}$ is calculated using  the law of propagation. 

From Eq.~(\ref{eq:BLS_fDMI}), $f_{\mathrm{DMI}}$ is proportional to the spin-wave wave vector $k$, which is selected by adjusting the incidence angle $\theta$. Choosing $\theta$ involves a trade-off: larger angles increase the frequency asymmetry (improving resolution), but reduce the optical signal-to-noise ratio.  A fixed incidence angle of \(\theta = 45^\circ\) represents an optimized experimental compromise for our setup: it achieves a wave vector \(k\) large enough to yield an \(f_{\text{DMI}}\) shift well above the experimental uncertainty and avoiding the sharp drop in the BLS scattering cross-section that occurs at higher angles, thus keeping the total measurement time manageable for systematic spatial mapping.

\subsection{Evaluation of D from BLS Data}

\begin{figure*}[!t]
    \centering
    \includegraphics[width=0.8\linewidth, page=1, trim={2cm 4cm 2cm 5cm}, clip]{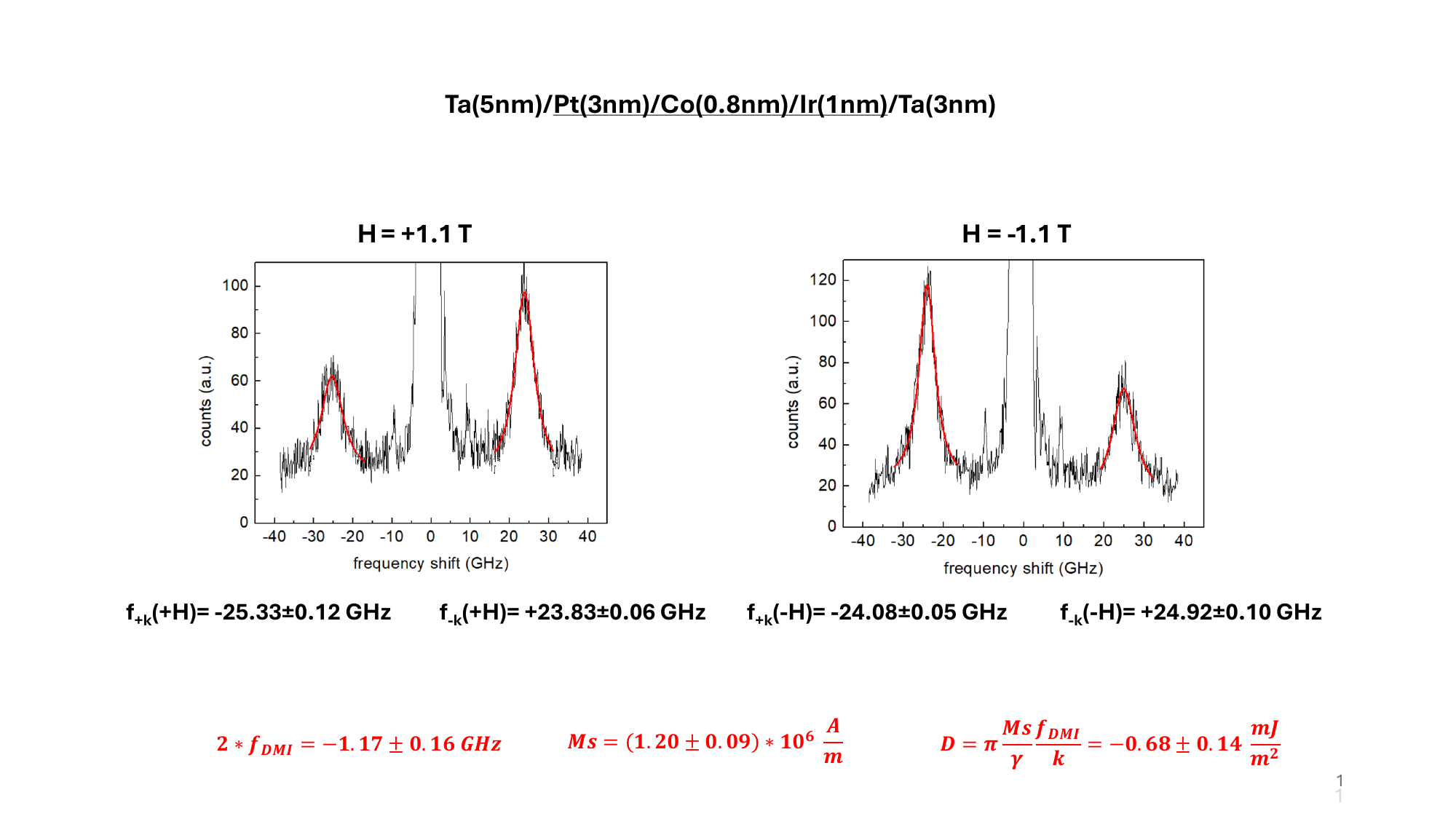}
    \caption{BLS spectra measured at a fixed incidence angle of $\theta=45^\circ$ for a Pt(3~nm)/Co(0.8~nm)/Ir(1~nm) sample under opposite applied magnetic field directions ($H=\pm1.1\text{ T}$). Red curves represent Lorentzian fits.}
    \label{fig:BLS_spectra}
\end{figure*}

Evaluating $D$ requires fitting the Stokes and anti-Stokes peaks (typically using Lorentzian functions) to extract the resonance frequencies $f_{\pm k}(\pm H)$. As an example, Fig.~\ref{fig:BLS_spectra} shows two BLS spectra measured on a Ta(5~nm)/Pt(3~nm)/Co(0.8~nm)/Ir(1~nm)/Ta(3~nm) (Pt/Co/Ir) multilayer sample.  This sample is part of the original series of round robin samples prepared at the University of Leeds\cite{magni2022}. Spectra were recorded for opposite magnetic field directions ($H = \pm 1.1\text{ T}$) at an incidence angle of $\theta = 45^\circ$.  From the fitted peak positions and Eq.~(\ref{eq:BLS_2spectra}), the DMI frequency and its combined standard uncertainty\footnote{ All measurement uncertainties reported throughout the manuscript refer to JCGM 100:2008 (GUM). The combined standard uncertainties are evaluated with a coverage factor k=1, corresponding to an approximate 68\% confidence interval.} is obtained, $f_{\mathrm{DMI}} = -0.59\pm0.08\text{ GHz}$.

To extract the iDMI constant $D$, Eq.~(\ref{eq:BLS_fDMI}) is inverted while propagating the uncertainties of all constituent parameters:
\begin{equation}
    D = \frac{\pi M_{\mathrm{s}}}{\gamma} \frac{f_{\mathrm{DMI}}}{k}
    \label{eq:BLS_Dfromf}
\end{equation}
The uncertainty in $k$ (set by the incidence angle) is typically around 1\%. The gyromagnetic ratio $\gamma$ can be measured with an accuracy of 2–3\% using ferromagnetic resonance (FMR) or field-dependent BLS measurements. Aside from $f_{\mathrm{DMI}}$, the largest uncertainty contribution usually stems from $M_{\mathrm{s}}$, which is commonly measured by VSM or SQUID magnetometry with an uncertainty of 5–10\%. As emphasized in \cite{magni2022}, accounting for dead layers or proximity effects is essential to accurately determine the effective magnetic volume and magnetization; neglecting these can introduce errors of 30–40\%. 

Laser heating must also be considered. Focused laser powers up to $200\text{ mW}$ can increase the local temperature in the probed region, altering temperature-dependent properties including $D$. On semiconductor substrates like Si, thermal dissipation limits this temperature rise to $\approx 20\text{--}30\text{ K}$ (where variations in $M_{\mathrm{s}}$ remain within experimental uncertainty), whereas insulating glass substrates can experience heating of several hundred degrees, risking sample damage.

Using the measured values  (with their respective standard uncertainty (type B)) $M_{\mathrm{s}} = 1.20 \pm 0.09\text{ MA/m}$ (corrected for proximity effects \cite{magni2022}), $\gamma = 194\pm4\text{ GHz/T}$, and $k = 16.7\pm0.2\text{ rad/}\mu\text{m}$, we obtain   $D = -0.68\pm0.11\text{ mJ/m}^2$. The relative combined standard uncertainty in \(D\) is derived by applying the root-sum-square method to the independent physical parameters, yielding a value of about \(\mathrm{15}\%\). Although individual Lorentzian peak positions can be fitted with low  standard uncertainty (1–5\% in Fig.~\ref{fig:BLS_spectra}), $f_{\mathrm{DMI}}$ depends on small frequency differences, expanding the relative error. Keeping the total number of parameters low is therefore critical to minimize uncertainty propagation.

\begin{figure*}[!t]
    \centering
    \includegraphics[width=0.7\textwidth, page=1, trim={6cm 3cm 6cm 1.5cm}, clip]{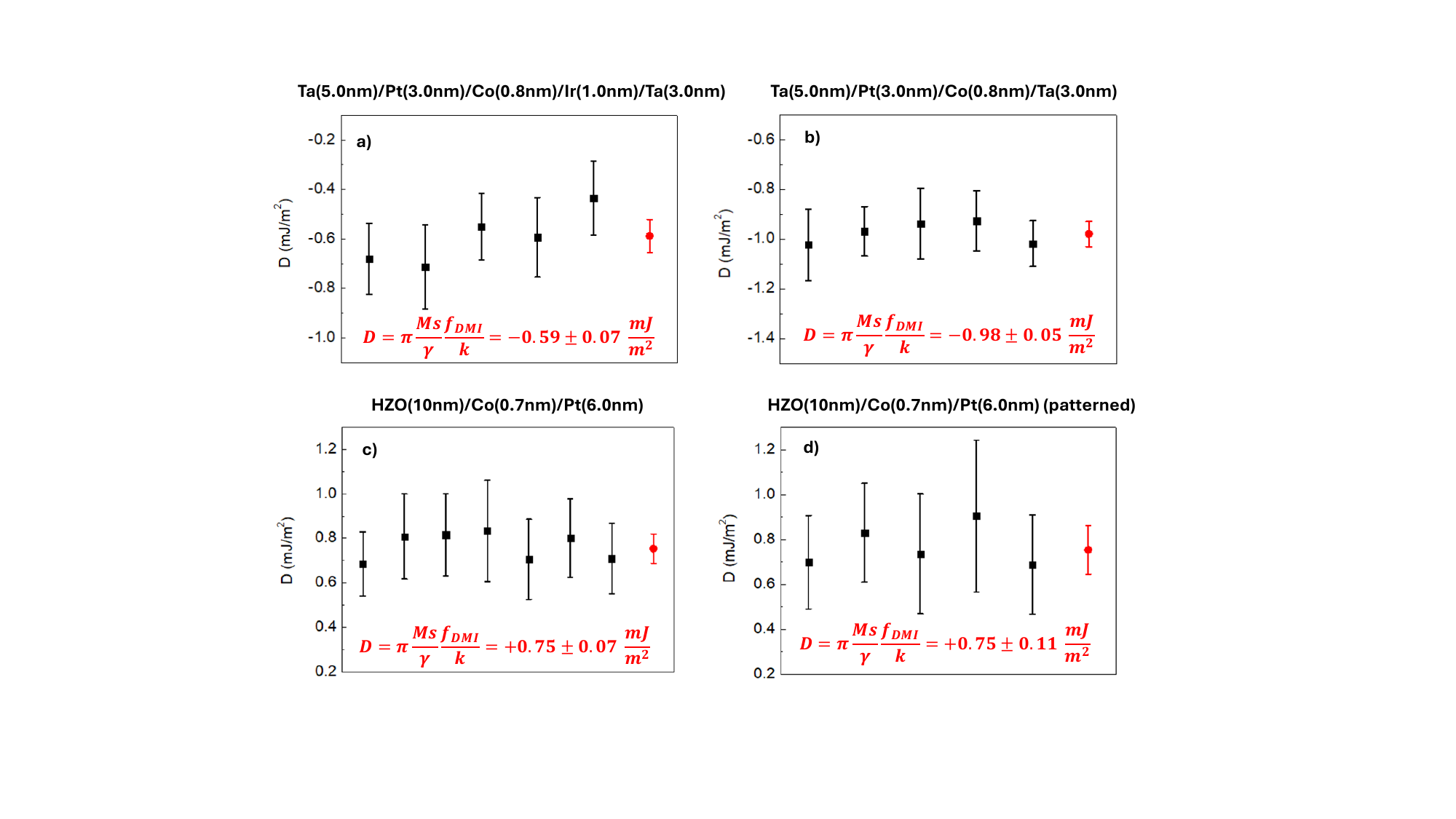}
    \caption{Measured $D$ values at various surface locations (black) and weighted averages (red) for (a) Pt/Co/Ir, (b) Pt/Co, (c) continuous as-grown HZO/Co/Pt, and (d) patterned HZO/Co/Pt pads.}
    \label{fig:BLS_statistics}
\end{figure*}

To improve statistical accuracy  -- an aspect that was not considered in the previous round-robin comparison -- measurements were repeated at different locations across the sample surface separated by several millimeters (Fig.~\ref{fig:BLS_statistics}). For the Pt/Co/Ir sample (Fig.~\ref{fig:BLS_statistics}a), spatial averaging reduced the relative uncertainty from $\approx 20\%$ for a single measurement to $\approx 10\%$ for the  relative combined standard uncertainty of the weighted mean. The black markers represent $D$ values obtained at five locations, while the red marker indicates the weighted average and its combined uncertainty. The spatial consistency confirms good overall sample uniformity despite minor local fluctuations.

An identical spatial analysis was performed on a Ta(5~nm)/Pt(3~nm)/Co(0.8~nm)/Ta(3~nm) (Pt/Co) sample, which lacks the top Ir layer (Fig.~\ref{fig:BLS_statistics}b). As expected \cite{Kuepferling2023}, omitting Ir increases the interfacial iDMI constant by roughly 50\% ($D = -0.98\pm0.05\text{ mJ/m}^2$). While single-point  combined standard uncertainties are comparable to the Pt/Co/Ir stack, the lower scatter across points indicates superior spatial uniformity, suggesting that the top Ir interface introduces additional spatial variation.

Finally, we investigated CMOS-compatible heterostructures consisting of Co/Pt deposited on a ferroelectric Hafnium Zirconium Oxide (HZO) layer: as-grown HZO(10~nm)/Co(0.7~nm)/Pt(6~nm)\footnote{ Nominal thicknesses are given in nm.}. Measurements were carried out both on continuous films (seven locations) and on $400\mu\text{m} \times 400\mu\text{m}$ square pads patterned by optical lithography and ion beam etching (five locations) (Fig.~\ref{fig:BLS_statistics}c,d). Spectra for the continuous film required 3–4 hours of integration per spectrum due to broader linewidths and lower signal-to-noise ratios, resulting in slightly higher individual  combined standard uncertainties, with a weighted average of $D = 0.75\pm0.07\text{ mJ/m}^2$. The patterned pads yielded a slightly higher relative combined standard uncertainty of 15-20\%, while maintaining the same weighted average $D$ value as the continuous film. Repeating measurements at multiple pads constrained the final relative uncertainty to 15\%. Notably, $D$ is positive in the HZO/Co/Pt stack because the Pt layer lies above the ferromagnetic Co layer, reversing the interface symmetry relative to the Pt/Co stacks.

\section{Determination of iDMI by Magneto-Optic Kerr Effect}

\subsection{Physical Principle: Creep Theory}

 The MOKE-based determination of iDMI exploits its influence on field-driven domain wall (DW) motion. In ultrathin magnetic films with perpendicular magnetic anisotropy (PMA), iDMI acts as an internal effective field, $H_D$, which favors chiral Néel walls over Bloch configurations. Measurements in the thermally activated creep regime \cite{lemerle1998} require low driving fields and long pulse durations, allowing DW motion to be resolved by standard magneto-optical imaging at room temperature \cite{magni2022, vanatka2015, hartmann2019, DiPietro2025}. Expanding magnetic bubble domains provide access to the angular velocity profile, $v(\theta)$, around the full bubble circumference, from which the iDMI constant $D$ can be extracted.
Under an out-of-plane driving field $H_z$ and a static in-plane (IP) field $H_x$, the creep velocity is described by
\begin{equation}
    v(\theta,H_z) = v_0 \exp\left[
    -\alpha_0\left(
    \frac{\chi(H_x,\theta)}{\mu_0 H_z}
    \right)^{1/4}
    \right] .
    \label{eq:creep_velocity}
\end{equation}
Here, $v_0$ and $\alpha_0$ are experimentally determined scaling parameters, while $\chi(H_x,\theta)$ is a dimensionless factor describing the IP-field and angular dependence of the activation barrier, normalized to unity at $H_x=0$. Different models of DW creep relate this factor to the equilibrium DW energy or to the DW stiffness.
The energy-based model introduced by Je et al.~\cite{je2013} associates this dependence with changes in the equilibrium DW energy density. For an IP field applied along $x$, the DW surface energy density is \cite{pakam2024}
\begin{equation}
\begin{aligned}
    \overline{\sigma}_{\mathrm{DW}}(H_{x}, \theta, \phi) &= 4\sqrt{AK_{\mathrm{eff}}} \\
    &\quad - \pi M_{\mathrm{s}}\Delta\mu_{0} \left[ H_{D}\cos(\theta-\phi) + H_{x}\cos(\phi) \right] \\
    &\quad + \frac{\ln(2)}{\pi}\delta\mu_{0}M_{\mathrm{s}}^{2}\cos^{2}(\theta-\phi) .
\end{aligned}
\label{eq:sigma_dw}
\end{equation}
$A$ is the exchange stiffness, $K_{\mathrm{eff}}$ is the effective anisotropy, $M_{\mathrm{s}}$ is the saturation magnetization, $\Delta =\sqrt{A/K_{\mathrm{eff}}}$ is the DW width, and $\delta$ is the magnetic film thickness. The angles $\theta$ and $\phi$ specify the DW normal and the internal DW magnetization, respectively, relative to the $x$ axis. The equilibrium DW energy density follows by minimizing over $\phi$:
\begin{equation}
    \overline{\mathcal{E}}_{\mathrm{DW}}(H_{x}, \theta) = \min_{\phi \in [0, 2\pi)} \overline{\sigma}_{\mathrm{DW}}(H_{x}, \theta, \phi) .
    \label{eq:energy_density}
\end{equation}
In this model, the factor entering Eq.~(\ref{eq:creep_velocity}) is
$\chi(H_x,\theta)=\overline{\mathcal{E}}_{\mathrm{DW}}(H_x,\theta)/
\overline{\mathcal{E}}_{\mathrm{DW}}(0,\theta)$.
Although this approximation describes creep velocities in some systems, its applicability is limited when the anisotropic elastic response of the DW becomes significant. We therefore compare this model with the stiffness-based description introduced by Pellegren, Lau, and Sokalski \cite{LauSokalski2017}
\begin{equation}
\tilde{\sigma}_{\mathrm{DW}}(H_x,\theta)
= \overline{\mathcal{E}}_{\mathrm{DW}}(H_x,\theta)
+ \sigma_{\theta\theta},
\label{eq:stiffness}
\end{equation}
where $\sigma_{\theta\theta}$ is the second partial derivative of the unminimized DW energy density with respect to $\theta$, taken at fixed internal angle $\phi$ and evaluated at the equilibrium angle that minimizes the DW energy (Eq.~\ref{eq:energy_density}). In this version of the model,
$\chi(H_x,\theta)=\tilde{\sigma}_{\mathrm{DW}}(H_x,\theta)/
\tilde{\sigma}_{\mathrm{DW}}(0,\theta)$ \cite{dipietro2026}.

\subsection{Measurement Protocol}

To measure asymmetric bubble expansion, a standard wide-field MOKE microscope is used, configured to apply simultaneous in-plane (IP) and out-of-plane (OOP) magnetic fields. In our setup, a commercial EVICO MOKE system was equipped with flat-pole electromagnet pieces capable of IP fields up to $150\text{ mT}$, alongside an OOP coil generating up to $100\text{ mT}$. Post-processing of polar Kerr images to extract DW velocities was performed using custom Python-based software \cite{DiPietro2025}.

To ensure the entire field of view remains in sharp focus, precise sample-stage tilt control is required. Strict planarity is critical: any angle misalignment causes the strong IP field to project a parasitic OOP field component onto the sample, which artificially biases the measured DW velocity.

The measurement sequence follows three standardized steps: saturation, nucleation, and expansion.
\begin{enumerate}
\item \textbf{Saturation:} The sample is magnetized to a uniform state using a high-amplitude $H_z$ pulse.
\item \textbf{Nucleation:} A initial domain (magnetic bubble) is nucleated using a square out-of-plane field pulse, $H_z^{\mathrm{nuc}}$. A reference Kerr image, $I(H_z^{\mathrm{nuc}})$, is captured to record the initial bubble size.
\item \textbf{Expansion:} A propagation field pulse, $H_z^{\mathrm{prop}}$, is applied to drive domain expansion. The expanded state is recorded in image $I(H_z^{\mathrm{prop}})$.
\end{enumerate}

\begin{figure}[!t]
    \centering
    \includegraphics[width=\linewidth]{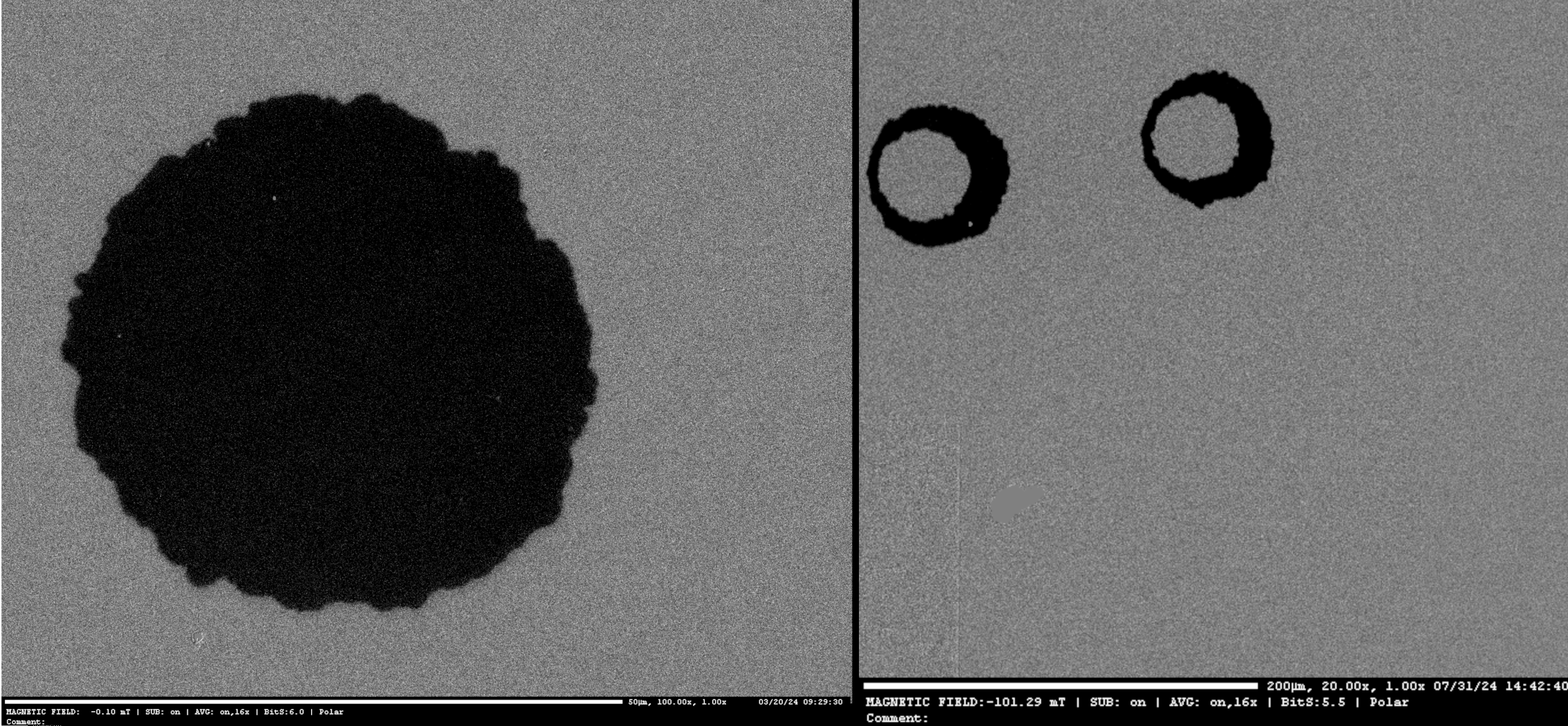}
    \caption{Nucleated magnetic bubble $I(H_z^{\mathrm{nuc}})$ imaged at $100\times$ magnification to highlight initial DW boundary roughness (left; scale bar: $50\text{ }\mu\text{m}$). Expanded domain difference image $I(H_z^{\mathrm{prop}}) - I(H_z^{\mathrm{nuc}})$ captured at $20\times$ magnification (right; scale bar: $200\text{ }\mu\text{m}$). Applied field pulse parameters: $H_z^{\mathrm{nuc}} = 27\text{ mT}$, $H_z^{\mathrm{prop}} = 20\text{ mT}$. Sample: HZO/Co/Pt (see Table~\ref{tab:HZO_params}).}
    \label{fig:MOKEbubble}
\end{figure}

Calculating the differential image $\Delta I = I(H_z^{\mathrm{prop}}) - I(H_z^{\mathrm{nuc}})$ subtracts background optical artifacts. As shown in Fig.~\ref{fig:MOKEbubble}, the initial bubble domain appears as a central light region, while the newly expanded area appears dark. This differential technique cleanly isolates the starting and ending DW boundaries, significantly improving spatial accuracy.

 Before data collection, the system is calibrated to ensure the planarity of the sample surface with respect to the IP field. 
The  DW velocity is obtained by the time duration of the field pulse and by a measure of the bubble displacement along the $x$ axis. This displacement $\Delta x$ is caused by the $H_z$ pulse produced by the OOP coil, under a (generally non-zero) IP field $H_x$. In the presence of a slight tilt $\alpha$ of the sample with respect to the $H_x$ direction, and additional $H_x \sin \alpha$ field is generated, accelerating or slowing the DW motion depending on its sign. A simple solution is the acquisition of two expanded bubble images under opposing $H_x$ field. By tuning a tilting stage holder the condition is reached in which $\Delta x_{+H_x} = \Delta x_{-H_x}$, within pixel resolution (at 50$\times$ magnification, $1 \,\mathrm{px} \approx 235 \, \mathrm{nm}$), confirming the elimination of stray OOP components.

After calibration, the creep scaling parameters $v_0$ and $\alpha_0$ are determined by measuring the DW velocity $v(H_z)$ at zero IP field across various OOP pulse amplitudes ($H_z$) and fitting the response to the creep power law. Pulses typically use a duration of $\Delta t = 100\text{ }\mu\text{s}$, with multiple pulses applied if larger displacements are required.

A known challenge in MOKE-based iDMI determination is that the domain expansion characteristics varies from bubble to bubble, which can increase measurement uncertainty. As shown in our recent work \cite{DiPietro2025}, this variation stems from a dependence of DW expansion velocity on the initial bubble domain radius. Small initial radii introduce a negative curvature-induced pressure (analogous to Laplace pressure in liquid droplets) that acts like a demagnetizing field, hindering domain expansion. To eliminate this systematic effect, a preliminary calibration is performed to identify the minimum radius, $R_b^{\mathrm{min}}$, above which velocity becomes independent of initial domain size. All subsequent measurements acquiring $v(\theta, H_x, H_z, R_b)$ are conducted for $R_b > R_b^{\mathrm{min}}$. The effective iDMI field $H_D$ is then extracted using either field-dependent velocity profiling or single-image anisotropic mapping.

\subsection{Evaluation of D from MOKE Data}

Once domain expansion images are acquired, the effective iDMI field $H_D$ (and the corresponding iDMI constant $D$) can be extracted through two primary approaches:

\subsubsection{IP-Field Dependent Velocity Profiling}

The most widespread method determines $H_D$ by measuring DW velocity $v$ as a function of an applied IP field $H_x$ swept along fixed angular directions (typically along the field axis, $\theta = 0$ and $\theta = \pi$). Fitting the resulting $v(H_x)$ velocity profiles against Eq.~(\ref{eq:creep_velocity}) yields $H_D$.

The primary strength of field-dependent profiling lies in its experimental robustness: $H_D$ is directly identified by the shift in the minimum of the $v(H_x)$ curve. In the absence of iDMI, domain wall velocity is symmetric around $H_x = 0$; an internal iDMI field shifts this velocity minimum to $H_{x,\mathrm{min}} = \pm H_D$, corresponding to the IP field that exactly compensates $H_D$. Because this approach relies directly on measured velocity-field curves, it minimizes dependence on auxiliary parameters and their associated uncertainties \cite{magni2022}. However, acquiring velocity profiles across many field points is time-consuming and less suited for rapid screening or constrained geometries.

\subsubsection{Single-Image Method}

\begin{figure}[!t]
    \centering
    \includegraphics[width=0.75\linewidth]{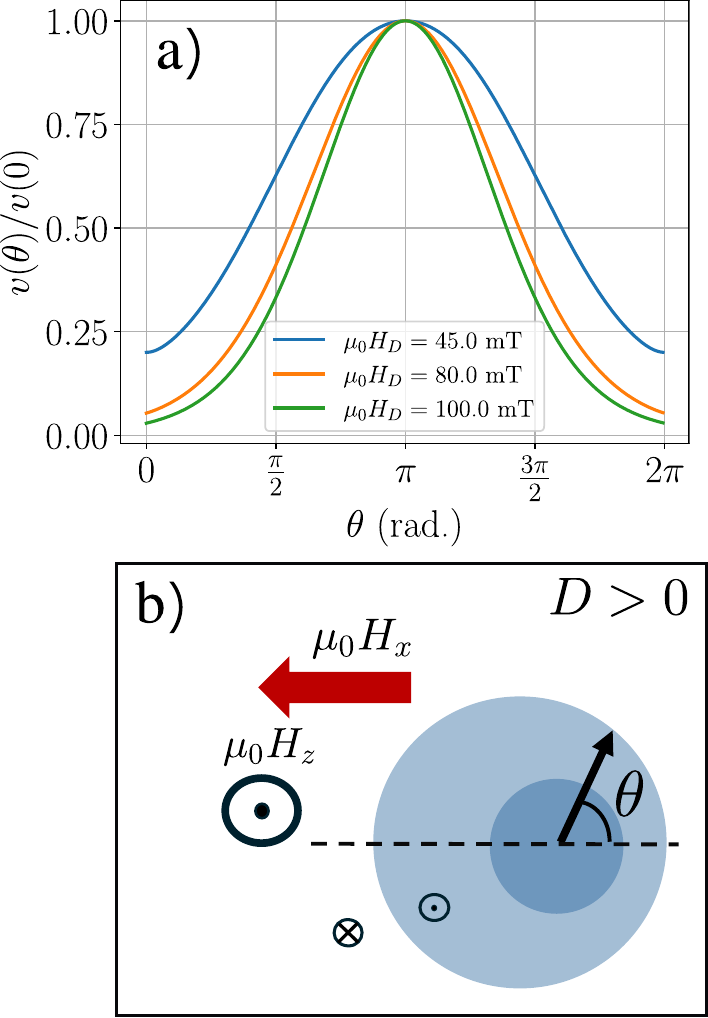}
    \caption{(a) Predicted anisotropic velocity profiles $v(\theta)$ for different values of $\mu_0 H_D$. (b) Schematic representation of asymmetric bubble expansion under simultaneous IP ($H_x$) and OOP ($H_z$) magnetic fields. The cartoon shows a positive DMI ($D > 0$) bubble expansion with an OOP applied field $H_z$.}
    \label{fig:aniso}
\end{figure}

Alternatively, $H_D$ can be extracted by mapping the full angular velocity profile $v(\theta)$ from a single asymmetric expansion event. Fitting this profile numerically to Eq.~(\ref{eq:creep_velocity}) yields $H_D$ (Fig.~\ref{fig:aniso}) \cite{pakam2024}.

The single-image method significantly increases throughput, making it well-suited for collecting large statistical datasets or characterizing spatially confined micro-patterned samples where domain expansion distance is constrained. However, because an analytical expression for the minimized energy density (Eq.~(\ref{eq:energy_density})) cannot be written in closed form, the fit relies on numerical minimization. This demands precise, independent measurement of all auxiliary material parameters prior to fitting, including saturation magnetization ($M_{\mathrm{s}}$), exchange stiffness ($A$), effective anisotropy ($K_{\mathrm{eff}}$), film thickness ($\delta$), and creep parameters ($v_0, \alpha_0$).

 The overall multiplicative prefactor $v_0$ can be eliminated as a fitting parameter by normalizing each angular velocity profile to its peak value. Care is required, however, because fluctuations in the normalization reference affect the entire profile and can influence the fitted $H_D$. We therefore normalize each profile by the arithmetic mean of the four measured velocities nearest the reference peak direction, which yields more stable $H_D$ estimates than normalization by a single measured value.

A fundamental challenge for both MOKE evaluation techniques is the absence of a single, universally accepted analytical model for chiral DW dynamics in the creep regime \cite{LauSokalski2017, hartmann2019}. Consequently, model selection must be carefully evaluated for each specific material system.

\subsubsection{Experimental Results}

A detailed analysis of asymmetric expansion by MOKE in the Pt/Co/Ir and Pt/Co round-robin samples using field-dependent velocity profiling was previously reported by the authors, noting that MOKE evaluations often yield systematically lower $D$ values than BLS on those systems \cite{magni2022}. Here, we focus instead on the HZO/Co/Pt sample stack and analyze the uncertainty of the high-throughput single-image method.

\begin{figure}[!t]
    \centering
    \includegraphics[width=0.8\linewidth]{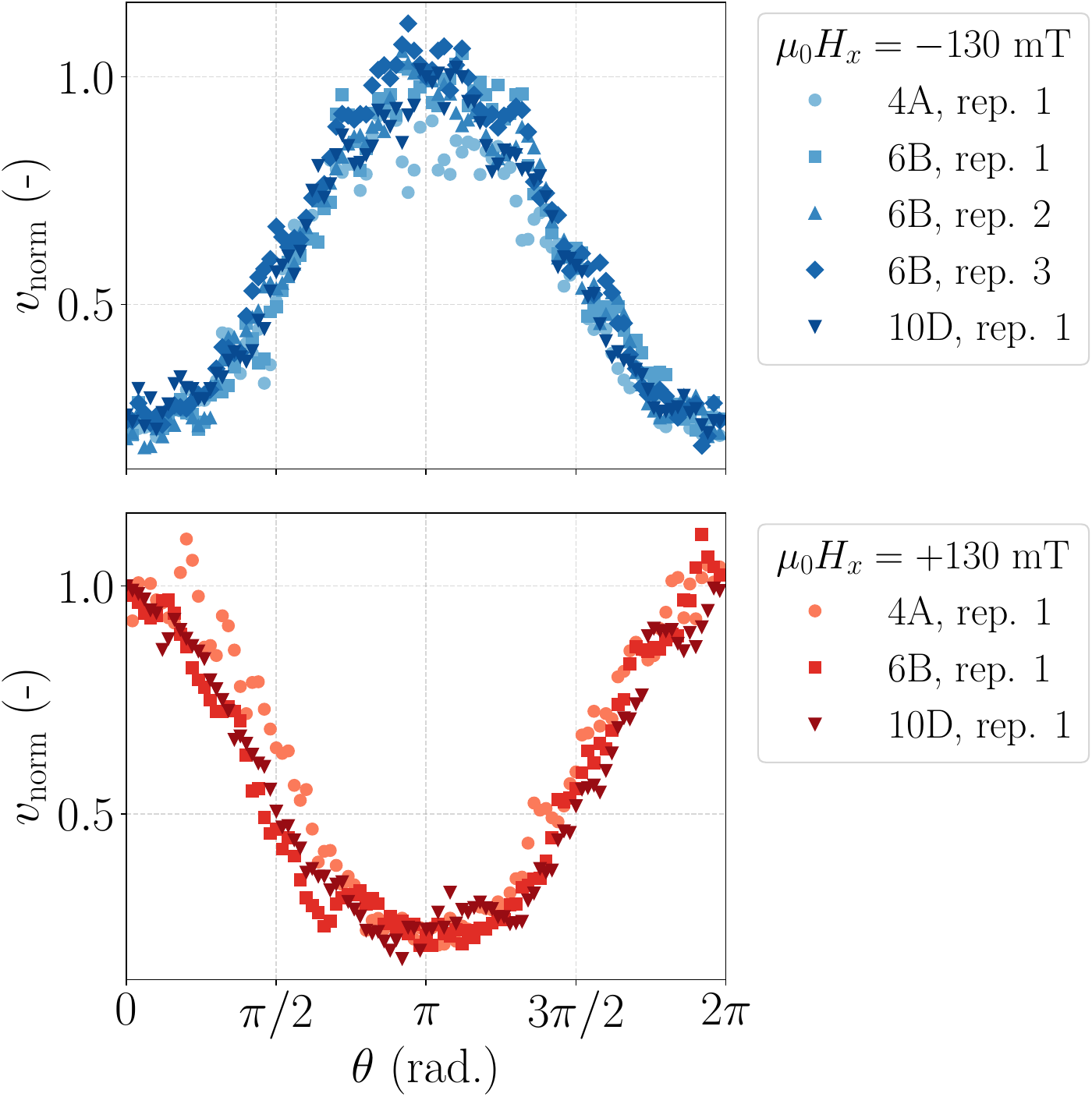}
    \caption{ Normalized Anisotropic velocity profiles $v_{\mathrm{norm}}$ measured across 8 distinct measurements performed on 3 pads of the as-grown HZO/Co/Pt sample. The label "rep." indicates an independent measurement on the same sample pad.}
    \label{fig:Capacitors_statistics}
\end{figure}

\begin{figure}[!t]
    \centering
    \includegraphics[width=\linewidth]{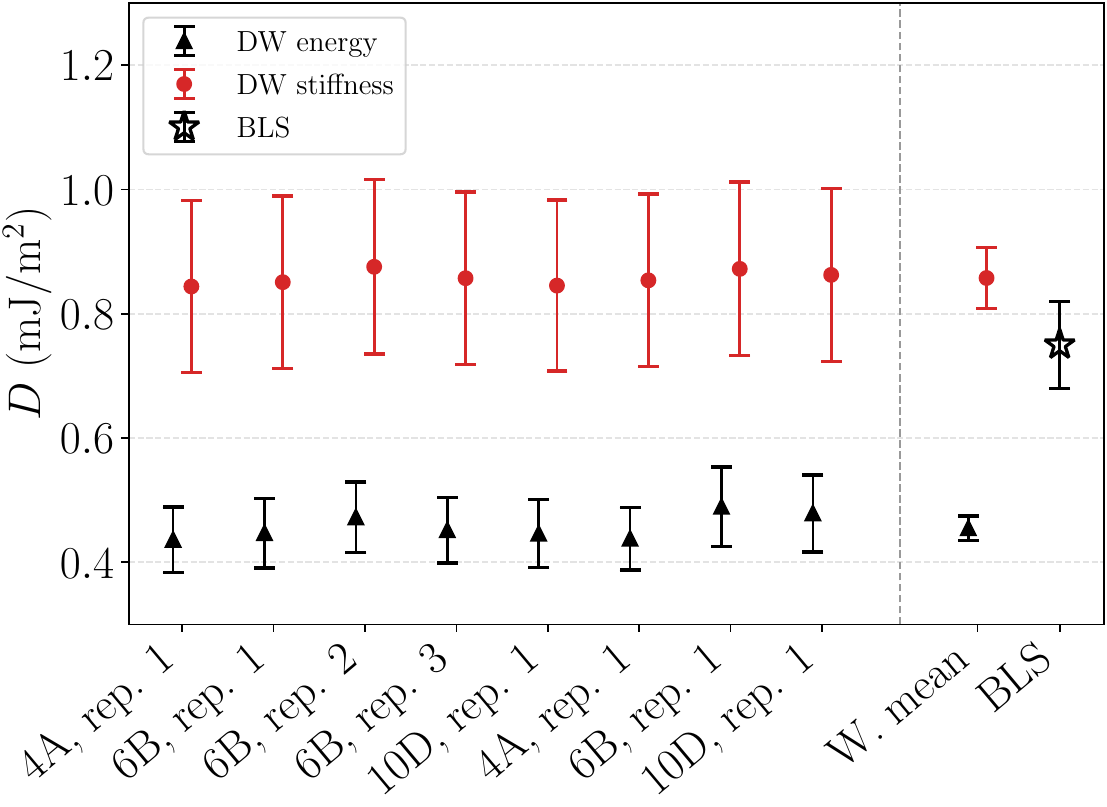}
    \caption{ iDMI constant values (D) obtained by fitting the measured $v(\theta)$ profiles shown in Fig.\ref{fig:Capacitors_statistics}. The black triangles are the values obtained by fitting the energy only model \cite{je2013,pakam2024} (Eq.\eqref{eq:energy_density}) and the red circles are the values obtained from the DW stiffness \cite{LauSokalski2017,dipietro2026} (Eq.\eqref{eq:stiffness}). The values separated by the dashed line are the average (weighted mean) values. The black star is the iDMI value obtained from BLS measurements. } 
    \label{fig:DMI_vals}
\end{figure}

 Eight measurements were performed on three different
$400\mu\text{m} \times 400\mu\text{m}$ patterned pads of the as-grown
HZO/Co/Pt stack, labeled 4A, 6B, and 10D in
Fig.~\ref{fig:Capacitors_statistics}. The anisotropic velocity profiles
\(v(\theta)\), recorded at \(\mu_{0}H_{x}=\pm130~\mathrm{mT}\) and
\(\mu_{0}H_{z}=19.4~\mathrm{mT}\), are shown in the same figure.
For each measurement, \(H_D\) was extracted by fitting
Eq.~(\ref{eq:creep_velocity}). We tested two creep models: one in which $\chi(H_x, \theta)$ depends on the domain wall energy density \cite{je2013,pakam2024} -- this is the model more commonly used in literature; and a second in which $\chi(H_x, \theta)$ depends on the domain wall stiffness \cite{LauSokalski2017,dipietro2026}. The creep
coefficient $\alpha_0$ was fixed at its independently determined value. The other auxiliary parameters used in the model are reported in Table~\ref{tab:HZO_params}. \(M_{\mathrm{s}}\), and \(K_{\mathrm{eff}}\) were were obtained from the independent magnetometry \cite{magni2022} and fixed at their nominal values. The Co thickness used in the model was obtained from the independent X-ray characterization. The exchange stiffness was not measured directly. We used a nominal value
of \(A=20~\mathrm{pJ\,m^{-1}}\) and assigned a uniform interval of
\(15\)-\(25~\mathrm{pJ\,m^{-1}}\), based on values reported for comparable ultrathin Co films~\cite{Zimmermann2019Comparison,
Bottcher2021Quantifying,Devolder2016Exchange}. This interval corresponds
to a standard uncertainty \(u(A)=5~\mathrm{pJ\,m^{-1}}\).
Uncertainties in \(A\), \(M_{\mathrm{s}}\), and \(K_{\mathrm{eff}}\)
were propagated using numerical sensitivity coefficients following
JCGM~100:2008, Sec.~5.1.3, Note~2. For each input
\(q\in\{A,M_{\mathrm{s}},K_{\mathrm{eff}}\}\), the complete
velocity-profile fit was repeated at \(q_0-u(q)\) and \(q_0+u(q)\), while
the other inputs were held at their nominal values. At each perturbed
value, \(D\) was recalculated as
\begin{equation}
D=\mu_{0}M_{\mathrm{s}}
\sqrt{\frac{A}{K_{\mathrm{eff}}}}\,H_D .
\end{equation}
The sensitivity coefficient and corresponding standard uncertainty
contribution were evaluated as
\begin{equation}
c_q \simeq
\frac{D[q_0+u(q)]-D[q_0-u(q)]}{2u(q)},
\qquad
u_q(D)=\lvert c_q\rvert u(q).
\end{equation} 
Assuming uncorrelated inputs, the combined standard uncertainty of each determination was
calculated as
\begin{equation}
u_{\mathrm{c}}(D)=
\sqrt{
u_{\mathrm{fit}}^{2}(D)
+u_A^{2}(D)
+u_{M_{\mathrm{s}}}^{2}(D)
+u_{K_{\mathrm{eff}}}^{2}(D)
},
\end{equation}
where \(u_{\mathrm{fit}}(D)\) is the standard uncertainty obtained from
the nominal fit. The individual $D$ estimates obtained from the fit and the above stated uncertainty estimation are show in Fig.\ref{fig:DMI_vals}. The individual determinations are compatible within their combined standard uncertainties. The overall $D$ value is calculated as their inverse-variance weighted mean, treating all measurements as independent. Its error bar represents the combined standard uncertainty of the weighted mean.

\begin{table}[!t]
\caption{Material Parameters for As-Grown HZO/Co/Pt.}
\label{tab:HZO_params}
\centering
\begin{tabular}{llll}
\hline
\textbf{Parameter} & \textbf{Symbol} & \textbf{Value} & \textbf{Unit} \\
\hline
Co Thickness             & $\delta$        & $0.68 \pm 0.03$ & nm \\
Saturation magnetization & $M_{\mathrm{s}}$ & $1.42 \pm  0.02$ & MA\,m$^{-1}$ \\
Effective anisotropy     & $K_{\mathrm{eff}}$ & $0.67 \pm 0.02$ & MJ\,m$^{-3}$ \\
Exchange stiffness       & $A$             & $20\pm5^*$          & pJ\,m$^{-1}$ \\
\hline
\multicolumn{4}{p{8cm}}{\footnotesize{\textit{All parameters, except $A$, were obtained from independent measurements with combined standard uncertainties (type B).}$^*$\textit{Literature value. \cite{Zimmermann2019Comparison,Devolder2016Exchange,Bottcher2021Quantifying}}}}
\end{tabular}
\end{table}

\section{Discussion}

The BLS method provides a reliable framework for determining the iDMI constant $D$, exploiting the non-reciprocal dispersion of Damon–Eshbach spin waves and modeling the response with a limited number of free parameters. This simplicity inherently minimizes the impact of error propagation compared to more complex models. An essential experimental protocol identified here is the implementation of field reversal, which is mandatory to decouple iDMI-induced non-reciprocity from inherent instrumental asymmetries within the interferometer. To address spatial inhomogeneities, we utilized multi-point spatial sampling across the sample surface. This practice significantly reduces statistical uncertainty through weighted averaging and serves as a diagnostic tool for verifying sample uniformity and interface quality. While BLS precision is sensitive to the signal-to-noise ratio and linewidth of the spectra, sometimes requiring acquisition times of 4–8 hours per spectrum, we demonstrate that rigorous statistical sampling can constrain uncertainty to 15–20\%, even in samples with low spectrum quality. 

 In contrast, creep-regime MOKE microscopy provides a faster and more accessible route to iDMI characterization, making the single-image method particularly suitable for high-throughput screening. Quantitative extraction nevertheless requires independent knowledge of parameters such as $M_{\mathrm{s}}$, $K_{\mathrm{eff}}$ and $A$ entering Eq.~(\ref{eq:sigma_dw}), some of which, particularly $A$, are difficult to determine accurately. The analysis must also account for the dependence of DW velocity on the bubble radius, which otherwise introduces a systematic bias in the extracted effective iDMI field.

 A more fundamental limitation is the absence of a universally accepted model of chiral DW dynamics in the creep regime \cite{LauSokalski2017, hartmann2019}. Although the energy-only model can adequately describe some material systems, including that studied by Pakam \emph{et al.}, for the present HZO/Co/Pt dataset it gives $D_{\mathrm{energy}}=0.46 \pm 0.02~\mathrm{mJ/m^2}$, approximately $38$\% below the BLS result of $0.75 \pm 0.07~\mathrm{mJ/m^2}$. Together with the measurement-to-measurement scatter in Fig.~\ref{fig:DMI_vals}, this suggests that the energy-only model does not capture all contributions to the velocity profiles. Including DW elasticity through the field-dependent DW stiffness \cite{dipietro2026} instead brings the MOKE-derived value ($D_{\mathrm{stiffness}}=0.86 \pm 0.05~\mathrm{mJ/m^2}$) into agreement with BLS.

 Since no independent criterion currently determines which creep model is appropriate for a given material system, model selection remains the principal limitation of quantitative MOKE characterization. MOKE and BLS can nevertheless be used complementarily: BLS can provide a reference $D$ value to constrain multi-parameter creep-model fits \cite{dipietro2026}, after which the faster MOKE method can screen similar samples or patterned structures. In this approach, however, the BLS result acts as a calibration constraint rather than an independent validation of the creep model.

\section{Conclusions}

In this work, we establish a comprehensive set of good practices for the reliable determination of the iDMI constant, $D$, using either BLS spectroscopy or MOKE microscopy. By investigating various heavy-metal/ferromagnet heterostructures, specifically Pt/Co/Ir, Pt/Co, and HZO/Co/Pt stacks, we identify primary sources of uncertainty and provide robust methodologies for their mitigation. Our analysis reveals that BLS inherently yields lower uncertainties due to the reduced number of independent parameters required to evaluate $D$. However, this technique remains highly sensitive to sample quality and may necessitate long measurement times to achieve sufficient signal-to-noise ratios. Conversely, MOKE microscopy offers the advantages of wider accessibility and high throughput, yet the determination of $D$ is complicated by a larger parameter space, including creep coefficients, which can propagate significant measurement error. Furthermore, the selection of an appropriate measurement model is critical and must be tailored to the specific magnetic properties of the sample. We recommend that researchers carefully weigh these experimental trade-offs to ensure a reliable determination of the iDMI constant. Adhering to the protocols established herein, incorporating field reversal symmetry, domain wall curvature corrections, and rigorous statistical sampling, is essential for achieving the reproducible iDMI values necessary to advance next-generation chiral spintronic devices.

\section*{Acknowledgment}
The work was supported by the MetroSpin project, funded by PRIN2022 SAYARY (CUP E53D2300183 0006) financed by the European Union - Next Generation EU and MUR (Ministero dell’Universita e della Ricerca). C.R. and G.G. acknowledge the project Spin-ORBit Effects in Two-dimensional magnets (SORBET) – ID 2022ZY8HJY - CUP D53D23002500006 financed by EU – Next Generation EU - “PNRR - M4C2, investimento 1.1 - “Fondo PRIN 2022”. We thank E. Darwin and A. Huxtable for the growth of the round robin multilayers at the University of Leeds.

\bibliographystyle{IEEEtran}
\bibliography{xampl}

\balance

\end{document}